\documentclass[
aps,
prapplied,
twocolumn,
english,
longbibliography,
superscriptaddress,
reprint,
bibnotes,
amsmath,
amssymb,
]{revtex4-2}

\usepackage[T1]{fontenc}
\usepackage{graphicx}   
\usepackage{dcolumn}    
\usepackage{bm}         
\usepackage{siunitx}    
\usepackage[colorlinks=true, linkcolor=black, urlcolor=black, citecolor=black]{hyperref}    

\DeclareUnicodeCharacter{03BC}{{$\mu$}}

\begin{document}

\title{Multiplexed cryo-CMOS control of an isolated double quantum dot}

\author{Mathieu~Darnas}
\affiliation{Univ. Grenoble Alpes, CNRS, Grenoble INP, Institut N\'eel, 38042 Grenoble, France}

\author{Mathilde~Ouvrier-Buffet}
\affiliation{Univ. Grenoble Alpes, CNRS, Grenoble INP, Institut N\'eel, 38042 Grenoble, France}
\affiliation{Univ. Grenoble Alpes, CEA, Leti, F-38000 Grenoble, France}

\author{Antoine~Faurie}
\affiliation{Univ. Grenoble Alpes, CEA, Leti, F-38000 Grenoble, France}

\author{Brian~Martinez}
\affiliation{Univ. Grenoble Alpes, CEA, Leti, F-38000 Grenoble, France}

\author{Jean-Baptiste~Casanova}
\affiliation{Univ. Grenoble Alpes, CEA, List, F-38000 Grenoble, France}

\author{Beno\^it~Bertrand}
\affiliation{Univ. Grenoble Alpes, CEA, Leti, F-38000 Grenoble, France}

\author{Candice~Thomas}
\affiliation{Univ. Grenoble Alpes, CEA, Leti, F-38000 Grenoble, France}

\author{Jean~Charbonnier}
\affiliation{Univ. Grenoble Alpes, CEA, Leti, F-38000 Grenoble, France}

\author{Jean-Philippe~Michel}
\affiliation{Univ. Grenoble Alpes, CEA, Leti, F-38000 Grenoble, France}

\author{Bruna~Cardoso~Paz}
\affiliation{Quobly, Grenoble, France}

\author{Yvain~Thonnart}
\affiliation{Univ. Grenoble Alpes, CEA, List, F-38000 Grenoble, France}

\author{Franck~Badets}
\affiliation{Univ. Grenoble Alpes, CEA, Leti, F-38000 Grenoble, France}

\author{Franck~Balestro}
\affiliation{Univ. Grenoble Alpes, CNRS, Grenoble INP, Institut N\'eel, 38042 Grenoble, France}

\author{Matias~Urdampilleta}
\affiliation{Univ. Grenoble Alpes, CNRS, Grenoble INP, Institut N\'eel, 38042 Grenoble, France}

\author{Tristan~Meunier}
\affiliation{Quobly, Grenoble, France}

\author{Baptiste~Jadot}
\email{baptiste.jadot@cea.fr}
\affiliation{Univ. Grenoble Alpes, CEA, Leti, F-38000 Grenoble, France}

\date{\today}

\begin{abstract}
Scalable spin-based quantum computing demands precise and stable control of a large number of gate-defined quantum dots while minimizing wiring complexity and thermal load. Control architectures based on sample-and-hold (SH) multiplexing techniques offer a promising solution by enabling sequential programming of several gate voltages using a limited number of input lines. However, the compatibility of such dynamic voltage refreshing with the stringent stability, noise, and speed requirements of quantum dot operation is an active subject of study. Here we experimentally demonstrate that a multiplexing cryo-CMOS circuit can reliably bias a silicon double quantum dot (DQD) at $\SI{0.5}{K}$. Exploiting the isolated regime, we show deterministic loading and isolation of four electrons and stable access to all five charge configurations from (4,0) to (0,4), despite the sequential voltage refreshing. We further demonstrate rapid voltage pulsing across an inter-dot transition, resolving single-electron tunneling events and stochastic switching at the \mbox{(1,3)--(0,4)} transition. These results confirm that SH-based multiplexed control is compatible with both static biasing and pulsing of isolated quantum dots, representing an important milestone toward scalable cryogenic control architectures for large-scale spin-qubit processors.
\end{abstract}
\maketitle

\section*{Introduction}
Spin qubits in gate-defined semiconductor quantum dots (QD) constitute a promising platform for scalable quantum computing thanks to their potential for dense integration, long coherence times and compatibility with standard CMOS fabrication \cite{philips_universal_2022, huang_high-fidelity_2024, takeda_quantum_2022, weinstein_universal_2023, chatterjee_semiconductor_2021}. Their structure, close to classical transistor, allows to envision a rapid scale-up following recent single- and two-qubit gate demonstrations in industry-compatible devices \cite{zwerver_qubits_2022, steinacker_industry-compatible_2025, hamonic_foundry-fabricated_2025}. However, they require individually-tuned biasing voltages to accommodate for qubit inhomogeneity \cite{thomas_rapid_2025, contamin_methodology_2022}. In large-scale fault-tolerant architectures requiring millions of qubits \cite{beverland_assessing_2022}, this large number of different signals presents a challenge in terms of number of wires from room temperature, heat load and crosstalk.

Cryogenic control circuits have therefore emerged as a key enabling technology to multiplex control signals and mitigate qubit inhomogeneity with minimal impact on the cryogenic environment. They are often based on a sample-and-hold (SH) approach, where individual gate voltages are stored in integrated capacitors, and then sequentially refreshed to compensate for leakage \cite{jadot_cryogenic_2023, bartee_spin-qubit_2025, enthoven_3v_2022, subramanian_scalable_2024, schreckenberg_sige_2023}. While this approach offers substantial scalability benefits, its compatibility with sensitive QD operation remains uncertain. Voltage drift, power dissipation near the qubits, charge injection during switching and limitation in pulsing speed are among the strongest concerns.

To date, direct experimental validation that SH-based multiplexed control can stably bias and rapidly manipulate an isolated quantum system has been limited either by the number of gates simultaneously controlled \cite{bartee_spin-qubit_2025} or by the lack of pulsing capability \cite{schreckenberg_sige_2023}. In this work, we address this question by interfacing a cryo-CMOS demultiplexer with a silicon double quantum dot (DQD) operated at $\SI{0.5}{K}$. Combining charge stability mapping and time-resolved pulsing, we demonstrate such a control scheme can maintain stable charge configurations while enabling fast dynamic transitions, thus validating its suitability for scalable quantum architectures.

\begin{figure}
    \centering
    \includegraphics[width=8.6cm]{"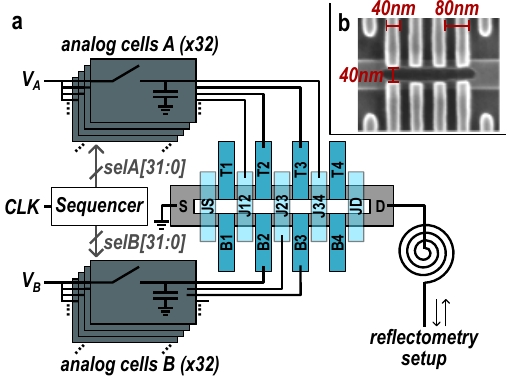"}
    \caption{\textbf{(a)} Schematics of the quantum device connections. The 7 innermost gates are connected to either $V_A$ or $V_B$ through the cryo-CMOS controller. The remaining gates are directly connected to room-temperature electronics. A \SI{500}{nH} Nb superconducting inductor is used for reflectometry readout. \textbf{(b)} Scanning Electron Microscope (SEM) picture of an identical quantum device before the J-gates are patterned.}
    \label{fig1}
\end{figure}

\begin{figure*}[!htbp]
    \centering
    \includegraphics[width=17.2cm]{"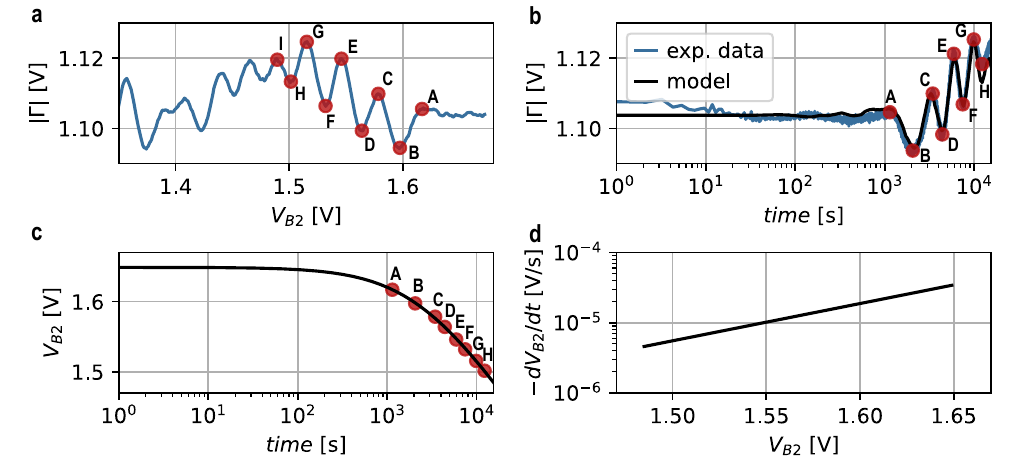"}
    \caption{Voltage drift of the analog cell reconstructed with Coulomb peaks. \textbf{(a)} Reference Coulomb peaks under gate B2. \textbf{(b)} Reflectometry signal $|\Gamma|$ recorded over $\SI{4.3}{hours}$ as $V_{B2}$ decreases from its initial value of $\SI{1.65}{V}$ towards $V_{DD} / 2 = \SI{0.9}{V}$. \textbf{(c)} Evolution of $V_{B2}$ reconstructed by matching the peaks from \textbf{(a)} and \textbf{(b)}. \textbf{(d)} Voltage drift as a function of the current voltage.}
    \label{fig2}
\end{figure*}


\section*{Experimental setup}

Figure \ref{fig1} presents a scheme of the experimental setup. A cryo-CMOS demultiplexing circuit \cite{jadot_cryogenic_2023} is connected via wire-bonding to a FDSOI DQD sample fabricated at CEA-Leti \cite{bertrand_tunnel_2023}, and placed on a test board anchored to a dilution refrigerator. Due to a $\SI{1}{mW}$ parasitic power consumption dissipated by unused functions within the cryo-CMOS controller \cite{jadot_cryogenic_2023}, the test board temperature is $\SI{0.5}{K}$, for an quantum device electronic temperature of $\SI{0.8}{K}$ in the isolated regime. The demultiplexer implements a sample-and-hold architecture in which multiple output voltages are sequentially updated and stored on integrated capacitors, allowing control of up to 64 gates using two analog voltage inputs.

The quantum chip consists of two parallel rows of four top gates (T1-T4 and B1-B4) separated by five coupling gates (JS, J12, J23, J34 and JD). The Si nanowire is cut longitudinally (Fig \ref{fig1}b) to suppress tunneling between the top and the bottom sides of the structure \cite{bertrand_tunnel_2023}, while maintaining a sufficient capacitive coupling between the top and bottom rows to perform charge sensing. Gates B2 and B3 form a DQD, probed by a sensing dot defined under gates T2 and T3. J23 controls the DQD inter-dot coupling. B1 and B4 are maintained at strong negative voltages to isolate the DQD from the electron reservoirs.

The seven innermost gates are connected to the cryo-CMOS circuit, while the outermost gates (T1, T4, B1, B4, JS and JD) are directly connected to room-temperature electronics, as they require a voltage range above $V_{DD}=\SI{1.8}{V}$. The voltages applied on these outside gates are kept constant except during the charge initialization protocol. For each data point of the following maps, a voltage refresh sequence is applied to set the required voltages on gates T2, B2, T3, B3, J12, J23 and J34 successively, alternating between the $V_A$ and $V_B$ voltage inputs to remove transient voltages. After this sequence, an unused output is selected (leaving all gates disconnected from room-temperature electronics), and a measurement point is taken. This refresh sequence lasts $\SI{6.4}{\micro s}$ to update all gate voltages, for an energy consumption of $\SI{28}{pJ}$ per refresh sequence (on top of the circuit's static power consumption).

The conductance through the sensing dot is probed using a RF-SET reflectometry technique, by connecting a $\SI{500}{nH}$ superconducting Nb inductor to the drain and using an amplification chain and a room temperature demodulation setup. We achieve a $\SI{91.5}{\%}$ charge detection fidelity between (0,4) and (1,3) states in $\SI{1}{ms}$, limited by the weak coupling between the sensor and the DQD.


\section*{Voltage drift}
Figure \ref{fig2} presents a measurement of the leakage of the analog cells, using the Coulomb peaks under gate B2 as a reference. First, we ramp $V_{B2}$ up to $\SI{1.65}{V}$ and we record the position of the peaks (Fig. \ref{fig2}a). All other gates controlled by the demultiplexer are kept at $\SI{0.9}{V}$ except for $V_{B3}=\SI{1.3}{V}$, in order to minimize their own voltage drift with respect to $V_{B2}$. Once the ramp is complete, the circuit is stopped on an unused output, leaving all innermost gates disconnected from room-temperature electronics. As the reflectometry signal $\Gamma$ is measured every second for the next $\SI{4.3}{hours}$ (Fig. \ref{fig2}b), we observe several peaks passing by as $V_{B2}$ decreases. Matching these peaks against the reference, we can reconstruct the drift of $V_{B2}$ (Fig. \ref{fig2}cd) which is on the order of $\SI{10}{\micro V/s}$ with an exponential dependence on the voltage. 

This voltage drift can be modeled by the sum of the integrated capacitor $C_s=\SI{70}{fF}$ and the parasitic capacitance $C_p\sim\SI{1}{pF}$ leaking through a MOS transistor. Compared to the results reported in \cite{jadot_cryogenic_2023}, the increase in parasitic capacitance brought by the connections to the quantum device leads to an improvement of the voltage drift by a factor $5$ at $\SI{1.5}{V}$. Conversely, we observe a reduction of bandwidth of the SH cell by one order of magnitude ($\SI{320}{MHz}$ to $\approx\SI{20}{MHz}$) due to the increase of parasitic capacitance. 

Typical voltage drifts reported in the literature vary between $\SI{0.3}{\micro V/s}$ \cite{bartee_spin-qubit_2025} to few $\SI{100}{\micro V/s}$ \cite{subramanian_scalable_2024} depending on the leakage mechanism and the choice of capacitance $C_s$. Taking our own voltage drift of $\SI{5.5}{\micro V/s}$ extracted at $\SI{1.5}{V}$, we could bias over $\SI{10}{M}$ independent outputs from one single voltage input and a $\SI{1}{MHz}$ external clock, for a maximum voltage error kept below $\SI{100}{\micro V}$ and a $\SI{3.54}{\micro W}$ power consumption. Because their leakage is orders of magnitudes smaller at cryogenic temperatures, SH cells are therefore able to bias large ensembles of qubits with a limited overhead in terms of wiring, operability and power consumption.

\section*{Charge manipulation in isolated regime}

\begin{figure}
    \centering
    \includegraphics[width=8.6cm]{"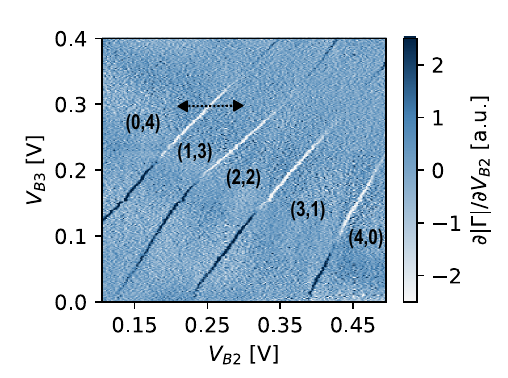"}
    \caption{Inter-dot charge transitions between B2 and B3 for a total number of electrons N=4. The dashed arrow indicates the region used for pulsing.}
    \label{fig3}
\end{figure}

\begin{figure*}
    \centering
    \includegraphics[width=17.2cm]{"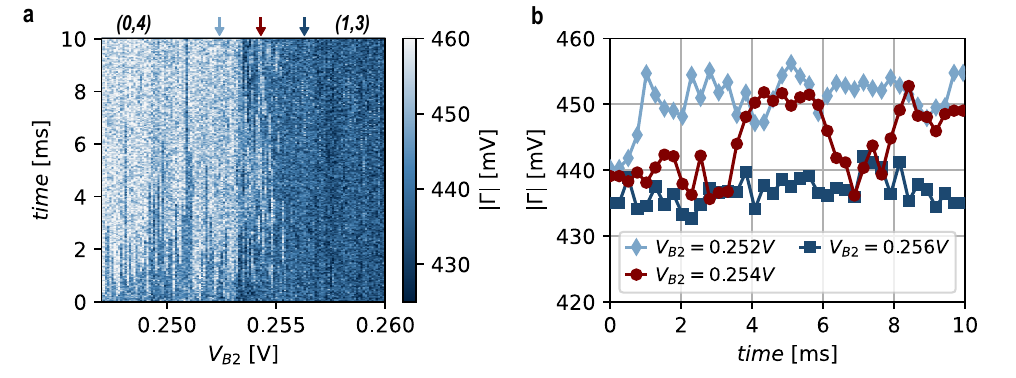"}
    \caption{\textbf{(a)} Single-shot readout traces obtained after pulsing from the (1,3) charge state to the (0,4)-(1,3) transition. \textbf{(b)} Traces showing three regimes: no tunneling, single tunneling or multiple tunneling events.}
    \label{fig4}
\end{figure*}

Using the charge isolation protocol described in Appendix A, we load 4 electrons into the B2-B3 double quantum dot and close the barrier gates B1 and B4 to prevent tunneling back to the electron reservoirs. Figure \ref{fig3} presents a charge stability diagram recorded by sweeping $V_{B2}$ and $V_{B3}$ with $V_{B1}=\SI{-1}{V}$ and $V_{J23}=\SI{1.3}{V}$. The RF-SET signal shows four parallel interdot transition lines, separating the five charge possible configurations (4,0), (3,1), (2,2), (1,3) and (0,4). Here the interest of the isolated regime is immediately apparent, as the fixed total number of charges greatly simplifies the readability of the charge stability diagram and reduces the noise on the charge detection.

Figure \ref{fig4}(a) presents a time-resolved measurement of the RF-SET following pulsing from the (1,3) charge state to an arbitrary $V_{B2}$ voltage. For $V_{B2} > \SI{0.255}{V}$, the system remains in the (1,3) state. For $V_{B2} < \SI{0.253}{V}$, a single tunneling event occurs after which the system remains in the (0,4) state. Pulsing near the transition, we observe multiple stochastic events, corresponding to fluctuations between the (1,3) and (0,4) charge states. Figure \ref{fig4}(b) shows single-shot traces corresponding to these three regimes. 

The well-defined charge transitions visible in Fig. \ref{fig3} indicate the ability of the cryo-CMOS demultiplexing circuit to maintain stable the voltages applied on B2 and B3 during the experiment, even if they are sequentially refreshed from the same input voltage before each measurement point. Furthermore, we observe in Fig. \ref{fig4} single-shot tunneling events, demonstrating that our cryo-CMOS circuit is able to apply detuning pulses with response time shorter than the inter-dot tunnel coupling ($\sim\SI{1}{ms}$ here).

\section*{Conclusions}
In this work, we demonstrated the control of an isolated double quantum dot by cryo-CMOS circuit using a sample-and-hold approach. The system maintains stable and well-defined charge configurations, while also supporting rapid voltage pulsing across interdot transitions, even though no two gates among B2, J23 and B3 are ever connected to room-temperature electronics at the same time. 

First, we use Coulomb peaks to monitor the leakage of our analog cells. We obtain a voltage drift of $\SI{5.5}{\micro V/s}$ at $\SI{1.5}{V}$, comparable to the state of the art. Extrapolating from this value, we could reasonably bias $\SI{10}{M}$ individual gates from one single input voltage and a $\SI{1}{MHz}$ clock. Then, we exploit this biasing capability by isolating four electrons into a double quantum dot and suppress tunneling towards the electron reservoirs. In this isolated regime, we observe four interdot charge transitions, demonstrating stable voltage control despite the sequential refresh scheme. Beyond DC biasing, we are also able to apply detuning pulses and we observe single-shot tunneling events.

This double capability to deliver a large number of independent DC voltages while allowing fast pulsing is a key requirement for scalable spin qubit architectures. Indeed, voltage detuning and interdot coupling need to be pulsed to at least three different operation points for single-qubit gate, two-qubit gate, and spin readout operations. Being able to include these baseband signals with a reduced wiring overhead is a step towards scalable large-scale spin qubit control.

While this work focuses on charge stability and tunneling dynamics, future studies must address the impact of multiplexed control on spin coherence and gate fidelities. Additionally, tighter co-integration of cryo-CMOS and quantum circuits, either monolithically or with 3D integration techniques, will introduce other challenges such as thermal management, crosstalk and footprint. Overall, this work marks a significant step toward practical, large-scale silicon-based quantum processors.

\section*{Acknowledgements}
This work is supported by the French National Research Agency under the program "France 2030" (PEPR PRESQUILE - ANR-22-PETQ-0002).
B. C. P. and T. M. acknowledge the support of EIC MCQube grant 101218381 and MCSquare grant 101136414.

The authors thank Edouard Deschaseaux, Meriem Guergour, C\'eline Feautrier, Fr\'ed\'eric Berger, Aurelia Plihon and Vincent Josselin for their technical input, as well as Heimanu Niebojewski, Pierre-Andr\'e Mortemousque and Emmanuel Chanrion for fruitful discussions.

\section*{Author contributions}
M.D. carried out the experiment with the help of M.O.B, M.U, and B.J.
B.B, C.T., J.C., J.P.M and Y.T. designed the quantum device, superconducting inductor and cryo-CMOS circuit.
A.F., B.M., J.B.C and B.C.P helped conceiving the experiment and analysing the experimental data.
B.J. supervised the project with help from F.Badets, F.Balestro and T.M.
B.J. wrote the manuscript with inputs from all the authors.


\section*{Data availability}
All data supporting the findings of this study are available from the corresponding author upon request.

\section*{Appendix A: Charge initialization}

\begin{figure}[!htbp]
    \centering
    \includegraphics[width=8.6cm]{"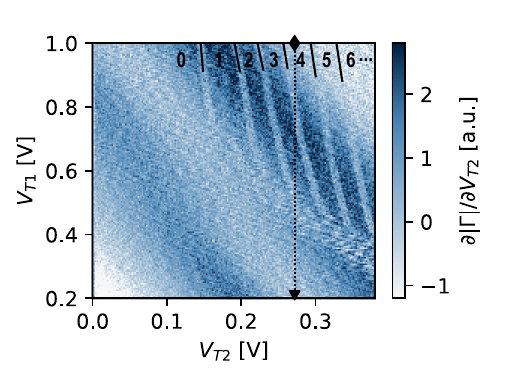"}
    \caption{Illustration of the charge isolation protocol for the top side of the structure. $V_{T1}$ is abruptly pulsed from $\SI{+1}{V}$ to $\SI{-1}{V}$, closing the coupling to the electron reservoir and isolating 4 electron into the (T2, T3) double quantum dot.}
    \label{fig_appendixA}
\end{figure}

Figure \ref{fig_appendixA} presents the initialization protocol. Sweeping the voltages applied on T1 and T2 and recording the RF-SET signal, we observe charge transitions attributed to a change in the finite number of electrons contained in the QD located under T2. $V_{T3}$ and $V_{J23}$ are kept at sufficiently low voltages to empty the quantum dot under gate T3 during the loading procedure. As T1 becomes less positive, the charge transition lines disappear, indicating suppression of charge tunnelling event to and from the electron reservoir, entering the so-called isolated regime. 

To initialize four electrons in the system, we pulse the system to ($V_{T1}=\SI{+1}{V}$, $V_{T2}=\SI{0.28}{V}$) and abruptly close T1 down to $V_{T1}=\SI{-1}{V}$. $V_{T3}$ and $V_{J23}$ can then be raised to allow charges to tunnel between the quantum dots defined under T2 and T3. In the data from the main text, we use the bottom side of the structure as the isolated double quantum dot, and the top side as the sensor. However, the charge isolation protocol is identical to the one presented in Fig. \ref{fig_appendixA} apart from the voltage $V_{B2}=\SI{0.24}{V}$ applied on B2 when B1 is closed.

\bibliography{PRApp25}

\end{document}